\documentclass{epl}

\title{Understanding the dynamics of segregation bands
\\of simulated granular material in a rotating drum}

\author{Nicolas Taberlet\inst{1,2} \and Wolfgang Losert\inst{2} \and Patrick Richard\inst{1}}
\institute{
  \inst{1} G.M.C.M., Universit\'e Rennes 1, UMR 6626, 35042 Rennes, France\\
  \inst{2} I.R.E.A.P., University of Maryland at College Park, College Park MD 20742, USA \\
}


\begin{document}
\maketitle

\begin{abstract}
Axial segregation of a binary mixture of grains in a rotating drum is studied using Molecular Dynamics (MD) simulations. A force scheme leading to a constant restitution coefficient is used and shows that axial segregation is possible between two species of grains made of identical material differing by size. Oscillatory motion of bands is investigated and the influence of the frictional properties elucidated. The mechanism of bands merging is explained using direct imaging of individual grains.

\end{abstract}

\section{Introduction}
Among the many puzzling phenomena exhibited by granular media, axial segregation~\cite{Oyama39} aka. banding is one of the least understood. Due to the fundamental interest as well as the numerous industrial applications~\cite{Jaeger96}, a great deal of both experimental~\cite{Zik94, Choo97, Hill97, Newey04, Khan04} and theoretical~\cite{Zik94, Savage93, Aranson99, Puri01} work has been devoted to the topic, but full understanding is still lacking. Molecular Dynamics simulation provides new insights in the understanding of the phenomenon since it allows one to vary all parameters and measure any physical property. The only large scale numerical study of axial segregation~\cite{Rapaport02} reports remarkable results but interactions between grains are derived from the Lennard-Jones potential, which is not well-suited for granular material. Our simulation uses a modified spring-dashpot force scheme leading to a restitution coefficient independent of the species of grains colliding. Here we show that axial segregation is possible between two species of grains made of identical material differing by size. We observe that a difference in the frictional properties of the two species of grain is not necessary to the onset of banding but does triggers oscillatory instabilities. Finally, the mechanism of bands merging is elucidated using direct imaging of individual grains. We also propose under which conditions coarsening may stop or slow significantly.

\section{Simulation Methods}

This letter presents results based on the Molecular Dynamics method (MD), a.k.a. Discrete Elements Method (DEM). This method deals with soft (but stiff) frictional spheres colliding with one another. Although not flawless, this method has been widely used in the past two decades and has proven to be very reliable. The drum is partially filled with a mixture of small and large beads (with respective radius $R_S = 2.4$ mm and $R_L = 2 R_S$). The total volumes of small and large grains are equal.
The number of grains varies from $5\times10^4$ to $5\times10^5$. The length of the drum, $l$, varies from $200\,R_S \,(\simeq 50$ cm) to $1000\,R_S \,(\simeq 250$ cm) and its radius is set to $40\,R_S \simeq 10$ cm. The density is the same for both kinds of beads: $\rho = 0.6 \, \mbox{g}/\mbox{cm}^3$. The rotation speed is set to $0.5 \, \mbox{rot}.\mbox{s}^{-1}$.

The force scheme used is a dashpot-spring model for the normal force, $F_{ij}^n$, and a regularized Coulomb solid friction law for the tangential force~\cite{Frenkel96}, $F_{ij}^t$: $ F_{ij}^n = k_{ij}^n \delta_{ij} -\gamma_{ij}^n  \dot{\delta}_{ij} , F_{ij}^t = min(\mu_{ij} F_{ij}^n, \gamma^t v^s_{ij})$  where $\delta_{ij}$ is the virtual overlap between the two particles in contact defined by: $\delta_{ij} = R_i + R_j - r_{ij}$. The force acts whenever $\delta_{ij}$ is positive and its frictional component is oriented in the opposite direction of the sliding velocity. $k_{ij}^n$ is a spring constant, $\gamma_{ij}^n$ a viscosity coefficient producing inelasticity, $\mu_{ij}$ a friction coefficient, $\gamma^t$ a regularization viscous parameter and $v^s_{ij}$ is the sliding velocity of the contact. If $k_{ij}^n$ and $\gamma_{ij}^n$ are constant, the restitution coefficient, $e$, depends on the species of the grains colliding. In order to keep $e$ constant the values of $k_{ij}^n$ and $\gamma_{ij}^n$ are normalized using the effective radius $R_{eff}$ defined by $1/R_{eff} = 1/R_i + 1/R_j$: $k_{ij}^n = k_0^n R_0 /R_{eff}$ and $\gamma_{ij}^n = \gamma_0^n R_{eff}^2/R_0^2$. The particle/wall collisions are treated in the same fashion as particle/particle collisions, but with one particle having infinite mass and radius. The friction coefficient $\mu_{\alpha \beta}$ can take five different values corresponding to
collisions between the small grains, the large ones, and the wall ($\alpha, \beta \in [S, L, W]$).
The following values are used: $R_0=4$ mm, $k_0^n=400$ N.$\mbox{m}^{-1}$, $\gamma_0^n=0.012\;\mbox{kg}.\mbox{s}^{-1}$ (leading to $e \simeq 0.9$), $\gamma^t=6\;\mbox{kg}.\mbox{s}^{-1}$ and $0.1 < \mu_{\alpha \beta} < 1$.

The equations of motion are integrated using the Verlet method using a time step $dt=1/30\,\Delta t$, where $\Delta t$ is the duration of a collision ($\Delta t \approx 10^(-3)$s). The simulations are typically run for $10^7$ time steps, corresponding to a few hundred rotations. The drum is filled by randomly pouring grains from above all the way along the drum axis, $z$. Rotation of the drum is started after the grains have settled. The drum is always less than half-filled (filling fraction between 20\% and 50\%) in order to speed up the dynamics.

\section{Axial Segregation}

Prior to discussing any results on axial segregation, let us mention that radial segregation~\cite{Cantelaube95} is observed throughout the length of the drum after only a few rotations. This type of segregation is of great importance for axial segregation since the latter can be seen as a fluctuation of the inner core, made of small grains in most cases. Some previous experimental work~\cite{Hill97, Newey04} showed that the inner core is still present in the final steady state as some others~\cite{Nakagawa97, Chicharro97} showed that the core disappears meaning that the bands are pure. Both situations were observed in our simulation, depending on the size of the drum, the filling fraction and the material properties.

\begin{figure}[htbp]
\begin{center}
\resizebox{14.4cm}{!}{\includegraphics*{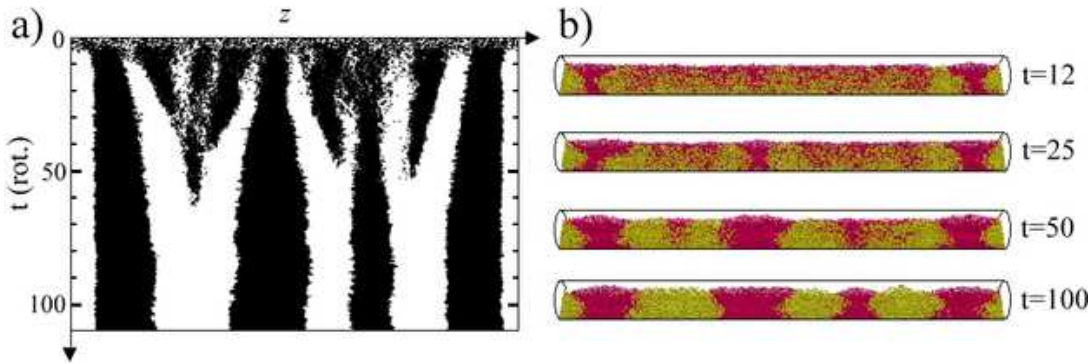}}
\caption{(in color online) a) Space-time plot for run A showing regions of high small-bead concentration. b) Snapshots of the drum during the coarsening of run A.}
\label{fig1}
\end{center}
\end{figure}

A wide range of parameters (drum length, radius, filling fraction, friction coefficients etc) was investigated and showed that our results are robust with respect to these parameters. In this letter, our goal is not to present a catalog of runs but to discuss the main features of the dynamics of axial segregation. Therefore, we would like to present only a limited number of runs which are in our opinion representative of many others.

The first run we would like to discuss consists of roughly 50,000 grains, with drum length, $l$=160 cm (run A). All the friction coefficients used for this run are equal ($\mu_{\alpha \beta}=0.5$). This means that the two species of grains made of identical material differ by size and not by frictional properties nor restitution coefficient. {Figure}~{\ref{fig1}a} is a space-time diagram showing the evolution of this system. The drum is virtually divided in thin cylindrical slices perpendicular to the rotation axis, each one corresponding to one pixel on {fig.}~{\ref{fig1}a}. The pixel is black if the concentration in small beads is higher than its average value (i.e. \#beads in slice $> N_S$/\#slices). The space-time plot shows a rich dynamics with bands appearing shortly after the rotation starts, with bands disappearing and merging with one another. The system seemingly reaches a steady state consisting of bands somewhat regularly spaced. The simulation was run for much longer than shown on {fig.}~{\ref{fig1}a} and no evolution was observed. However, it is well-known that coarsening can take place over several thousands of drum rotations. Let us mention here that the dynamics changes drastically when the seed of the random function is changed. The evolution shows the same features (banding, merging...) but can lead to a different meta-stable state. {Figure}~{\ref{fig1}b} shows snapshots of the drum at different times during the coarsening. The birth and merging of bands can be clearly seen. The last snapshot corresponds to the final steady state and shows 5 bands of large grains (L-bands), and 4 of small grains (S-bands). A plot of the concentration in small beads shows that in this final state, the bands are pure, i.e. the radial core has disappeared. Let us mention that the bands next to the walls are always S-bands in our simulations.

The first conclusion that can be drawn is that segregation by size only (with constant density $\rho$) is possible.
Since the micromechanical properties of grains may depend on their size, it is very difficult to experimentally study segregation by size only. As discussed below, a difference in frictional (or collisional) properties can affect the dynamics of segregation. Yet, our simulations show that such a difference is not necessary for axial segregation.

\begin{figure}[htbp]
\begin{center}
\resizebox{7.2cm}{!}{\includegraphics*{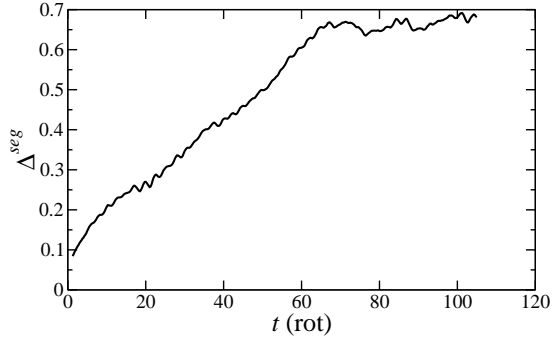}}
\caption{Degree of segregation, $\Delta^{seg}$, vs. time, for run A.}
\label{fig2}
\end{center}
\end{figure}

Let us now define a segregation function with the aim of quantifying the degree of axial segregation, regardless of the radial segregation. Our definition is similar to that of~\cite{Shoichi98} :
$$ \Delta^{seg}(t) = \displaystyle \frac{1}{l} \int_0^l \frac{|\,c_s(z,t)-\overline{c_s}\,|}{\overline{c_s}} \; dz $$ 
where $c_s(z,t)$ is the local concentration in small beads along the axis (i.e. the number of small beads per unit of length), and $\overline{c_s} \equiv N_s/l$ the average value of $c_s(z,t)$. When the two types of grains are mixed together, the concentration of small beads is uniform in the $z$ direction, meaning that $\Delta^{seg}(t)=0$, whereas if the grains are segregated in any number of pure bands (i.e. no radial core), $\Delta^{seg}(t)=1$.
{Figure}~{\ref{fig2}} is a plot of $\Delta^{seg}(t)$. The degree of segregation increases steadily until it reaches a plateau when two L-bands merge into one (i.e. disappearance of one S-band). The transition between the growth and the plateau is very sharp, meaning that the dynamics freezes. This shows that if further coarsening exists, it must occur on a much longer time-scale.

\section{Oscillations}

In this section we will discuss the sideways oscillatory motion of bands described several times in the literature(~\cite{Choo97,Khan04} with binary mixtures, ~\cite{Newey04} with ternary mixtures) but never explained. Such oscillations are observed in a wide range of drum lengths and radii but a drum shorter than that of run A is used here with the aim of saving computation time. Another advantage of a shorter drum is that there are fewer bands interacting with one another, making it easier to study their dynamics. The results presented here remain valid for longer drums. Two runs are presented in this section: run B and C. Both systems consist of 13,000 grains in a drum whose length is $l=40$ cm. In run B, all friction coefficients are identical ($\mu_{\alpha \beta}=0.6$) whereas in run C they differ ($\mu_{SS}=0.3, \, \mu_{LL}=0.6 ,  \, \mu_{SW}=\mu_{LW}=0.5 \mbox{ and } \mu_{SL}=(\mu_{SS}+\mu_{LL})/2)$.

\begin{figure}[htbp]
\begin{center}
\resizebox{14.4cm}{!}{\includegraphics*{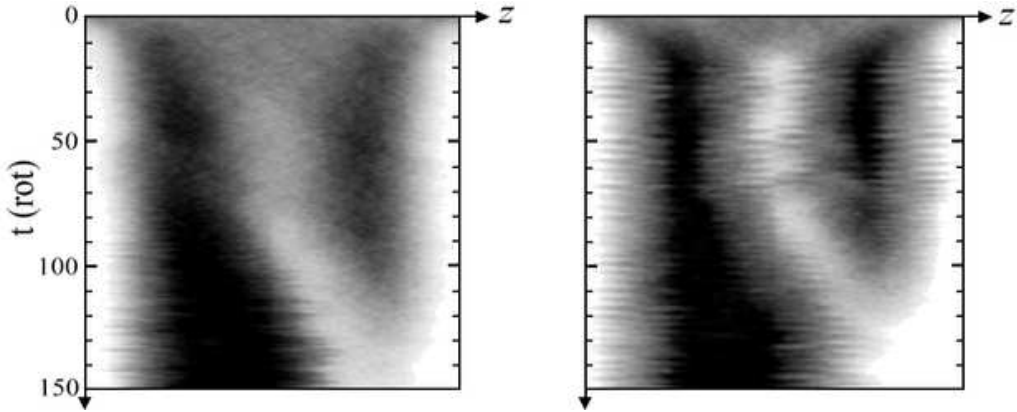}}
\caption{Space-time plot of $c_s(z,t)$ for runs B(left) and C(right).}
\label{fig3}
\end{center}
\end{figure}

{Figure}~{\ref{fig1}a} is a binary space-time plot: black and white represent regions of high small-bead and large-bead concentration respectively. To allow for a finer study of the dynamics one can plot the space-time evolution of the concentration $c_s(z,t)$ itself (using a gray scale).
{Figure}~{\ref{fig3}} shows the space-time plots of runs B and C. The dynamics of the two runs are similar although not identical. They both start with the birth of 3 L-bands and 2 S-bands. After a transient, the 2 S-bands merge into one, leading to 2 L-bands and 1 S-band. We would like to emphasize that except for the friction coefficients, the parameters of runs B and C are identical in every point, including the random seed. This explains why they exhibit the same general features.

\begin{figure}[htbp]
\begin{center}
\resizebox{14.4cm}{!}{\includegraphics*{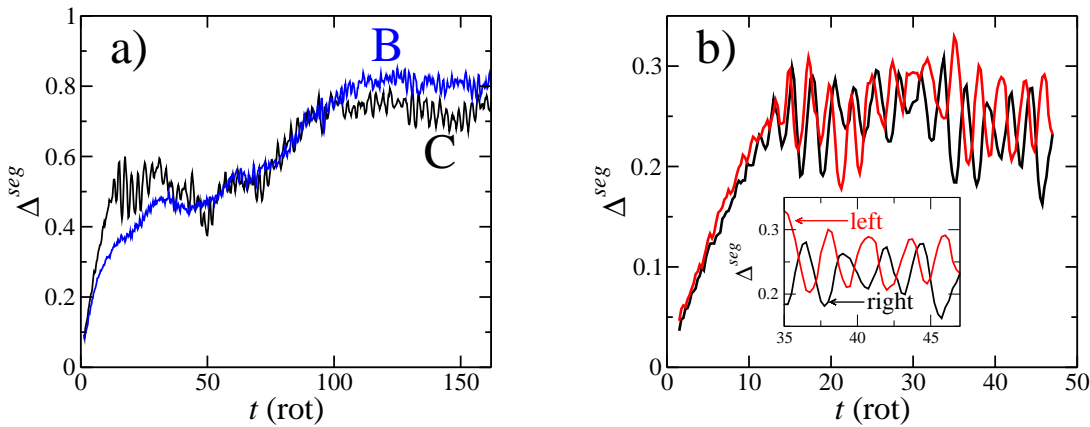}}
\caption{a) Segregation function $\Delta^{seg}$ vs. time for runs B and C. b) Partial segregation functions $\Delta^{seg}_{left}$ and $\Delta^{seg}_{right}$ vs. time for run C.}
\label{fig4}
\end{center}
\end{figure}

\noindent However, there exists a striking difference between the two runs: the space-time plot of run B seems to be smoother than that of run C. The fluctuations visible on run C are not noise but correspond to actual oscillations of the bands. This can be confirmed by performing a 2D-Fourier transform of {fig.}~{\ref{fig3}}: there exists a neat peak for run C and none for run B. We find the position of a L-band to oscillate as the width remains constant, whereas the position of a S-band remains constant as its width oscillates (as described in~\cite{Newey04}). Moreover, two neighboring bands of a kind (i.e. separated by a band of the other kind) are out of phase (see {fig.}~{\ref{fig4}b}). The absence of oscillations in run B shows that the oscillatory motion originates from the difference of frictional properties between species: when $\mu_{\alpha \beta}$ is independent of the species, the oscillations disappear.

{Figure}~{\ref{fig4}a} is a plot of the segregation function for runs B and C. It shows a two-step dynamics: after a rapid increase the function reaches a first plateau which is succeeded by a second plateau due to the merging event. Moreover, the oscillations mentioned above are clearly visible.
In order to study the oscillations of only one S-band at a time, one can define a partial segregation function $\Delta^{seg}_{left}(t)$ using the same definition as $\Delta^{seg}(t)$ but integrating from $0$ to only $l/2$ ($\Delta^{seg}_{right}(t)$  being the integral on the [$l/2, l$] interval). {Figure}~{\ref{fig4}b} presents the evolution of $\Delta^{seg}_{left}(t)$ and $\Delta^{seg}_{right}(t)$ for run C prior merging and shows very clear oscillations. The inset is a closer view of these functions just before the merging of the two S-bands. One can see that the oscillations of the two S-bands are out of phase, which as been observed experimentally~\cite{Newey04}.

\section{Merging}
In this last section we would like to discuss the mechanisms of merging. When a band disappears, what becomes of its grains ? Are they distributed to the neighboring bands or do they merge with only one of their neighbors ? The simulation is a powerful tool to address these questions. Indeed, most experiments focus only on the free surface of the media, which is not satisfactory. It is experimentally possible (but difficult) to measure the subsurface concentrations using MRI~\cite{Hill97} or optical techniques~\cite{Khan04}. It is however impossible to identify the original location of a grain (i.e. no particle tracking is possible). For the same reason, the space-time plots are of little help in understanding the mechanisms of merging. Moreover, the apparent mass (or width) of a band on a space-time plot is not a reliable source of information. For instance, the mass of the final S-band on {fig.}~{\ref{fig3}} (for both runs B and C) seems to be much larger than the total mass of the bands it originates from, which is absurd.

\begin{figure}[htbp]
\begin{center}
\resizebox{14.4cm}{!}{\includegraphics*{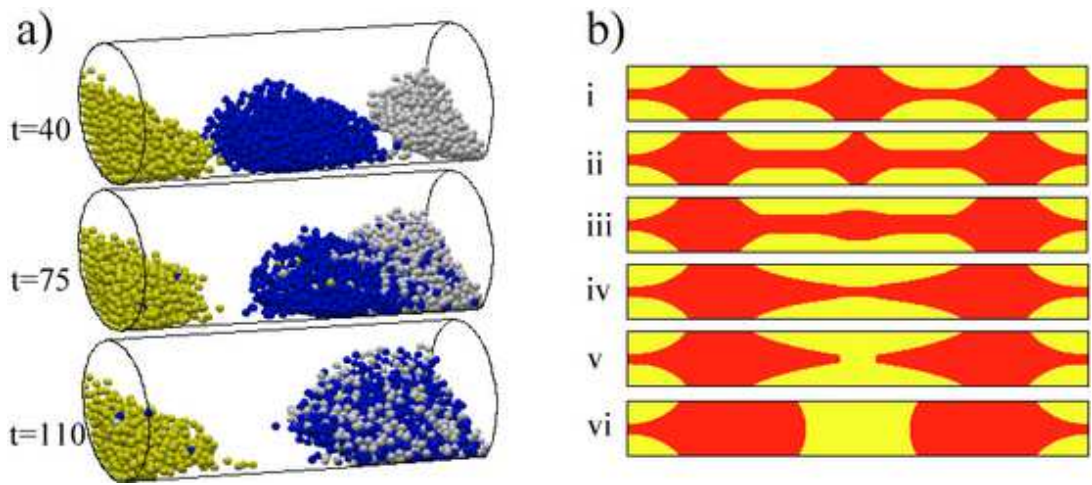}}
\caption{(in color online) a) Snapshots of the large beads in run C. The color of each grain is chosen according to the position at $t$=40 rot. b) Schematic cross sections showing a typical merging event.}
\label{fig5}
\end{center}
\end{figure}

{Figure}~{\ref{fig5}a} presents three snapshots of run C, before, during and after merging (respectively $t=$ 40, 75 and 110 rotations). For clarity, the small grains are not shown on these snapshots, but would fill in the voids left by the L-bands. Note that radial segregation still exists at time $t=$ 40 rot. (see {fig.}~{\ref{fig3}}), i.e. a core of small beads runs through the middle L-band. The color of each grain is chosen according to its position at time $t=$ 40 rot. (yellow for the left band, blue for the middle one and gray for the right one). The origin of each grain is then visible throughout the duration of the run. The composition of the two final L-bands is surprising. The left L-band in the final state is made of almost exclusively yellow beads whereas the right L-band is a well-mixed combination of blue and gray beads. More than 99\% of the middle (blue) particles choose the right side. Conversely, less than 1\% of the yellow particles are dragged to the right. The initial middle and right L-bands merge together as the left one is completely ignored.

This tendency for a L-band to merge with only one of its neighbors holds in longer systems (such as run A). In run C, there exist only 2 S-bands before merging. Therefore, the outcome of the merging for those is clear: the 2 bands must mix. Using longer drums (as in run A), we found that during a merging event, a S-band splits and feeds both the left and right neighboring S-bands. This is of course possible only if the S-band is not located next to a wall (like for run C). To summarize, a merging event consists of the loss of one band of each kind, but with two distinct mechanisms. Two neighboring L-bands collapse, ignoring all the others as one S-band splits and feeds its neighbors on both sides. 

{Figure}~{\ref{fig5}b} is a schematic view of a typical merging event, showing a cross section of the granular media. In the initial state (i), there are 3 S-bands connected through a radial core. As the middle S-band shrinks its grains are shared between the two other S-bands (ii). >From the large particles standpoint, this means that the two central L-bands merge together (iii). The left and right S-bands form insurmountable obstacles for the large grains. The evolution can either stop in the state (iv), where radial segregation still exists or continue and lead to a state where radial segregation has disappeared. Let us mention that as suggested by recent experiments~\cite{Flaten04}, the radial core could possibly reform but this was never the case in our simulations.
In run A, the death of the radial core coincides with the sudden arrest of the evolution ({fig.}~{\ref{fig2}}). This suggests that the exchanges between S-bands are possible only if the latter are connected through a radial core. Whenever this communication line is broken, the S-bands cannot interact and the evolution stops.

\section{Conclusion}

In this letter granular axial segregation is numerically studied using 3D molecular dynamics simulations. We show that this type of segregation can occur in a mixture of two species of grains differing by size only (i.e. equal densities, frictional properties, Young modulus etc). 
Although a difference in frictional properties between the two species is not necessary to observe axial segregation, it leads to the onset of oscillations in the band position or width. This suggests that the oscillations observed in experiments may originate from a difference in the frictional properties between species of grains.
The mechanisms of band merging are elucidated by tracking the positions of individual grains during a merging event.
They consists of two complementary events. On one hand, the disappearance of one S-band : its grains are shared between the two neighboring S-bands through the radial core, on the other hand, the collapse of two L-bands : the two L-bands surrounding the decaying S-band mix together and are oblivious to the other L-bands.
Finally, we introduce a segregation function that measures the degree of axial segregation in the medium. This function is a powerful tool to study the dynamics of banding. It allows for precise measurement of the period of band oscillations and shows that the coarsening process can stop or slow dramatically when the radial core breaks.


\acknowledgments
The authors acknowledge support from NASA grant NAG32736.

\end{document}